\documentclass[12pt,a4paper]{article}
\title{Vortex-like Structures in the Skyrme~-~Einstein Chiral Model}
\author{Yu.~P.~Rybakov,
         A.~M.~Tarabay,
          and I.~G.~Chugunov
(Department of Theoretical Physics,              
	Peoples' Friendship University of Russia, Moscow, Russia)}

\begin{document}
\maketitle

 \abstract
     {A regular method is suggested for constructing vortex-like solutions with
cylindrical symmetry in the Skyrme~-~Einstein chiral model. The method is
based on the expansion of metric and field functions in power series with
respect to the two small parameters entering the model. The length mass
density of the vortex is estimated.
 }                                               
\section{Introduction}

The Skyrme chiral model [1] proved its efficiency in modelling the structure
of baryons and light nuclei [2] via soliton-like configurations endowed with
the topological charge (winding number), the latter being interpreted as the
baryonic one. Recent years the interest to general relativistic description of
Skyrmions arose in view of possible astrophysical applications of topological
solitons coupled to gravity [3-5]. The economy of the Skyrme~-~Einstein
 model in treating nuclear and gravitational interactions of matter being
its very attractive feature, one should however indicate the nonintegrability
of the equations of motion that forced the authors of the aforementioned
papers to use numerical methods. In lieu of keeping up with this tradition
we elaborate in the present work an analytic method of solving  the equations
for vortex configurations with the aim to approximate three-dimensional
structures by closed vortices and to expand, in future, this method to the
configurations with spherical or axial symmetry.

\section{Equations of the Skyrme~-~Einstein model: neutral vortex case}

The Lagrangian density of matter in the model in question can be constructed
from the chiral current $l_\mu=U^+\partial_\mu U,\,\,\mu=0,1,2,3,\,\,U\in
SU(2),$ and reads
\begin{equation}
{\mathcal L}_{m}=-\frac{1}{4\lambda^{2}}{\rm
Sp}\,(l_{\mu}l^{\mu})+\frac{\epsilon^2}{16}{\rm
Sp}([l_\mu,l_\nu][l^\mu,l^\nu]), \end{equation}
where $\epsilon, \, \lambda$ are constant parameters. For the configurations
with cylindrical symmetry the metric is chosen in the form
\begin{equation}
ds^2=e^{2\mu}dt^2-e^{2\alpha}d\xi^2-e^{2\beta}d\varphi^2-e^{2\gamma}dz^2,
\end{equation}
where the functions $\mu,\,\alpha,\, \beta,\, \gamma$ depend on the radial
coordinate $\xi \in (-\infty,+\infty)$. As a first step we consider the simplest
case of the neutral vortex, when the chiral matrix $U$ is taken in the form
\begin{equation}
U=\exp(i\tau\Theta(\xi)),\quad
\tau=\left[ \begin{array}{ccc}
              0 & e^{-i\varphi}\\
   e^{i\varphi} & 0
\end{array}
\right].
\end{equation}

Putting (3) into (1) and taking into account of the Einsteinian gravity
Lagrangian density $ {\mathcal L}_{g} = R/2\ae $, where $R$ is the scalar curvature
and $\ae$ is the Einstein gravitational constant, one finds the following
expression for the action functional:

\begin{eqnarray}
\lefteqn {{\mathcal A}  =  2\pi\int\!\! dt \!\!\int\!\! dz\!\! \int\!\! d\xi
e^{\mu+\gamma} \left[\frac{1}{\ae}e^{\beta-\alpha} (\beta'\gamma'
+\mu'\beta'+{} \right. } \nonumber  \\ & & {} + \mu'\gamma') -
  \frac{1}{2\lambda^{2}}\left(e^{\beta - \alpha} \Theta'^{2} +
e^{\alpha-\beta}\sin^{2}\Theta \right) -{} \nonumber \\ & &\left.{}
 -\epsilon^{2}e^{-(\alpha+\beta)}\Theta'^{2}\sin^{2}\Theta \right].
\end{eqnarray}

 One infers from (4) that the theory admits the discrete symmetry
$\mu \rightleftharpoons\gamma$. Therefore making the substitution
$\mu = \gamma$ and imposing the harmonic coordinate condition [6]
\begin{equation}
\alpha =
\beta + 2\gamma,
\end{equation}
one transforms the functional (4) as follows:
\begin{eqnarray}
\lefteqn {{\mathcal A}  =  2\pi\!\!\int\!\! dt\!\! \int\!\! dz\!\!\int\!\! d\xi \!
\left[\frac{1}{16\ae}(u'^{2} - v'^{2}) -
\frac{1}{2\lambda^{2}}\Theta'^{2} \times {} \right.}
          \nonumber\\
  & &\left.{} \times   (1 + 2\epsilon^{2}
\lambda^{2} e^{-\frac{v}{2}}\sin^{2}\Theta) - \frac{1}{2\lambda^{2}}
e^{u-v}\sin^{2}\Theta \right]\!\!,
\end{eqnarray}
where $u= 4(\beta + \gamma),\;\, v=4\beta. $

The functional (6) admits the evident scale symmetry:
\begin{equation}
\xi \rightarrow \sigma\xi, \,\, z \rightarrow \sigma z, \,\,
u \rightarrow u - 2 \ln \sigma,\,\, v \rightarrow v,
\end{equation}
that will be used later for simplifying the asymptotic behavior of
solutions at space infinity.

Now we introduce the dimensionless parameters in the problem by scale
dilation:
\begin{displaymath}
\xi \rightarrow \sigma x, \,\, z \rightarrow \sigma z, \,\,
v \rightarrow v + 2 \ln \sigma^{2},\,\, u \rightarrow u +\ln\sigma^{2},
\end{displaymath}
where $\sigma^{2} = 2\lambda^{2}.$ Then we come to the quasimechanical
problem with $x$ playing the role of "time" and with the Hamiltonian "action"
(measured in the units $2\pi/\lambda^{2}$) of the form
\begin{eqnarray}
\lefteqn { S = \frac{1}{2} \int dx \left[\frac{1}{\nu}(u'^{2} - v'^{2}) - {}
\right.}
                 \nonumber  \\
& & \left.{} - \Theta'^{2}(1 + 2\epsilon^{2} e^{-v/2} \sin^{2}\Theta) -
e^{u-v} \sin^{2}\Theta \right],
\end{eqnarray}
where $\nu= 8\ae/\lambda^{2}.$ If one uses the natural units $\hbar=c=1$ and
the experimental value of the parameter $\lambda=2/F_{\pi},\,\,F_{\pi} =
186\,\, \mbox{MeV},$ then one gets $\nu\approx 1.26 \times 10^{-38}.$ The other
dimensionless constant $\epsilon$ appearing in the functional (8) takes the
experimental value [7] $\epsilon\approx 0.13.$ Thus there exist two small
parameters in the problem: $\nu\sim 10^{-38}$ and $\epsilon^{2}\sim 10^{-2},$
that allows one to utilize the perturbation scheme.

First of all we write down the variational equations corresponding to (8):

\begin{equation}
\frac{1}{\nu}u'' =  -\frac{1}{2}e^{u-v} \sin^{2}\Theta,
\end{equation}
\begin{equation}
\frac{1}{\nu}v'' =  -\frac{1}{2}e^{u-v} \sin^{2}\Theta
-\frac{\epsilon^{2}}{4}e^{-v/2}\Theta'^{2}\sin^{2}\Theta,
\end{equation}
\vspace{-9 pt}
\begin{eqnarray}
\Theta''(1 + \epsilon^{2} e^{-v/2}\sin^{2}\Theta ) -
\frac{\epsilon^{2}}{2}e^{-v/2}v'\Theta'\sin^{2}\Theta  = & {\,} & {}
        \nonumber \\
{} =\quad \frac{1}{2}\sin2\Theta(e^{u-v}-
\epsilon^{2}e^{-v/2}\Theta'^{2}),&{\,} &{}
\end{eqnarray}
and the  ``energy'' integral
\begin{equation} \frac{1}{\nu}(u'^{2} -  v'^{2}) + e^{u-v}
\sin^{2}\Theta = \Theta'^{2}(1 + \epsilon^{2} e^{-\frac{v}{2}}
\sin^{2}\Theta).
\end{equation}

 Using the scale symmetry (7) we impose the regularity conditions at space
infinity:
\begin{equation}
x  \rightarrow +\infty,\, v\rightarrow  +\infty, \,u-v \rightarrow 0,\,
\Theta \rightarrow{\mathcal O}(e^{-x}),
\end{equation}
and on the axis [6]:
\begin{eqnarray}
x  \rightarrow -\infty,\quad v\rightarrow  -\infty,\quad \Theta \rightarrow
\pi,{} & &{}
        \nonumber \\
\mid u-v\mid < \infty ,\qquad v'\exp\left(\frac{v-u}{2} \right) \rightarrow 4.&&
\end{eqnarray}
As follows from the equations (9) - (12), the regularity conditions (13) and
 (14) imply the asymptotic behavior of solutions at space infinity:
\begin{equation}
v=u=4x ,\,\,\Theta = A e^{-x},\,\, A ={\rm const},
\end{equation}
and on the symmetry axis:
\begin{equation}
v=v_{a}+4Dx,\,\,u=u_{a}+4Dx,\,\, \Theta= \pi -
 Be^{Dx},
\end{equation}
with the integration constants
$
v_{a},\,u_{a},\,D,\,B
$
 satisfying the
constraint
 \begin{equation}
 D\exp\left(\frac{v_{a}-u_{a}}{2}\right) = 1.
 \end{equation}

\section{Structure of vortex-like solutions}

 Now we take into account of the mechanical analogy and rewrite the Eq. (12)
in the form of Hamilton~-~Jacobi equation
 \begin{eqnarray}
 S_{\Theta}^{2} & = & \left(1 +
\epsilon^{2} e^{-\frac{v}{2}} \sin^{2} \Theta \right) \times {}
        \nonumber \\
 & & {} \times \left[ \nu (S_{u}^{2} - S_{v}^{2}) +
 e^{u - v} \sin^{2}\Theta \right].
\end{eqnarray}
Due to small
  value of the parameter $\nu$ we can represent the solution to the Eq. (18)
 as the power series expansion \begin{equation} S = \frac{1}{\nu}F + G + \nu
 H + \cdots.  \end{equation} Inserting (19) into (18) and equating the terms
 with the same powers in $\nu$ order by order, we get the equations for the
 unknown functions $F, G, H,\cdots.$ The equations for the function $F$ read
 \begin{equation}
 F_{\Theta} = 0,\quad  (F_{u} - F_{v})(F_{u} + F_{v}) = 0.
 \end{equation}
 The asymptotic behavior (15) being equivalent to the relations (as $x
 \rightarrow\infty$):
 \begin{displaymath}
 F_{u} = u' = v' = -F_{v},
 \end{displaymath}
 one concludes that the general solution to (20) turns out to be
 \begin{equation}
 F = F(u - v).
 \end{equation}

 The form (21) of the function $F$ implies the following equation satisfied
 by the function $G$:
  \begin{eqnarray}
 \lefteqn{ G_{\Theta}^{2}  =
  (1 + \epsilon^{2} e^{-\frac{v}{2}} \sin^{2}\Theta) \times  }
	   \nonumber\\
    & & \times \left[ 2 F'(u-v)(G_{u} + G_{v})+ e^{u-v} \sin^{2} \Theta\right].
  \end{eqnarray}
 The variables in the Eq. (22) can be separated by the substitution
 \begin{eqnarray} F(u-v) & = & 8 e^{(u-v)/2}, \\
G & = & e^{(u-v)/2} \sum_{n=0}^{\infty} \epsilon^{2n}e^{-nv/2}
 V_{n}(\Theta),
 \end{eqnarray}
 where the choice (23) is motivated by the
 nonuniform term in (22) and by the asymptotic behavior (15). Inserting
 (23) and (24) in (22) one obtains the recurrence relations satisfied by the
 functions $V_{n}(\Theta)$:

\begin{eqnarray}
\samepage
\sum_{k=0}^{n} V'_{k} V'_{n-k} + 4
 (n-1) \sin^{2} \Theta V_{n-1}\quad = {} & &{}
    \nonumber \\
{} = \quad \delta_{n0}\sin^{2}\Theta + \delta_{n1}
 \sin^{4}\Theta\, . & &{}
 \end{eqnarray}
  For $n = 0$  one derives from (25) and (15) that
 \begin{equation}
 V_{0} = 1 - \cos\Theta.
 \end{equation}
 For $n = 1$ one gets, in view of (26), the equation
 \begin{equation}
 2\sin\Theta V_{1}' + 4V_{1} = \sin^{4}\Theta .
 \end{equation}

  Taking into account of the relations (15) and (16), one concludes that
 the solution $V_{1}(\Theta)$ to the Eq. (27) has the two branches
 $V_{1}^{\pm}(\Theta)$ with the domains
 \begin{displaymath}
   D^{+} = \{0\le\Theta\le\frac{\pi}{2}\},\; D^{-} =
 \{\frac{\pi}{2}\le\Theta\le\pi\}.
 \end{displaymath}
 The corresponding functions attain the forms:

 \begin{eqnarray*}
\arraycolsep = 2pt
 V_{1}^{-}  =  - \frac{1}{6} (1+\cos\Theta)^{2}\left(\frac{4}{1-\cos\theta} + 3
-\cos\Theta\right), 
\end{eqnarray*}
\vspace{-9 pt} 
\begin{eqnarray}
 V_{1}^{+}  =  \frac{1}{6}(1-\cos\Theta)\sin^{2}\Theta.  
\end{eqnarray}
For $n\ge 2 $ one obtains the following
integral representation for the solution to the Eq. (25):  
\begin{eqnarray}
\arraycolsep = 2pt
\lefteqn{V_{n}^{\pm} = -\left(\frac{1+\cos\Theta}{1-\cos\Theta}\right)^{n}\! \int
\limits_{a_{\pm}}^{\Theta}\! \frac
{d\Theta}{\sin\Theta}\left(\frac{1-\cos\Theta}{1 + \cos\Theta}\right)^{n}
\times }
 \nonumber \\
&& {} \times \left(\frac{1}{2} \sum_{k=1}^{n-1}V_{k}'V_{n-k}'+
2(n-1)\sin^{2}\Theta V_{n-1}\right),
\end{eqnarray}
where $ a_{+}=0,\,\,a_{-}=\pi$.In particular, 
for $n=2$ from (28) and (29) one deduces
that

\begin{eqnarray}
\arraycolsep=2pt
\lefteqn{ V_{2}^{\pm}  =   \left(\frac{1+\cos\Theta}{1-\cos\Theta}\right)^{2}
\left(-\frac{8}{9} \ln\left((1\pm\cos\Theta)/2 \right)  
+ A_{0}\delta^{\pm} {} \right. }
      \nonumber \\ 
& &  \mp \frac{4}{9}\tan^{\pm2}\frac{\Theta}{2} \left.{} 
+ \sum_{n=1}^{5} A_{n}^{\pm}(1 - \cos\Theta)^{n}\right),
\end{eqnarray}
where the following numerical coeficients are introduced:
\begin{eqnarray*}
& &A_{0} = -\frac{22}{135},\,\,A_{1}^{\pm} = \pm\frac{2}{3},\,\,
A_{2}^{\pm} = \frac{2}{9},\,\,A_{3}^{\pm} = \frac{5}{54},\\
& & A_{4}^{\pm} = \frac{1}{24},\,\; A_{5}^{\pm} = -\frac{1}{40},\,\;\delta^{\pm}=(1 \mp 1)/2.
\end{eqnarray*}

Within the scope of the first two terms in the expansion (19) the following
normal system of equations emerges:
\begin{displaymath}
\frac{1}{\nu}u'  =S_{u}=\frac{1}{2} e^{\frac{(u-v)}{2}} \left(\frac{8}{\nu}+
\sum_{n=0}^{\infty}\epsilon^{2n} e^{-\frac{nv}{2}} V_{n}(\Theta)\right),
\end{displaymath}
\begin{eqnarray}
\frac {1}{\nu} v' \quad =\quad - S_{v}\quad =\quad \frac{1}{2}
e^{(u-v)/2} \left(\; \frac{8}{\nu} \quad +{}\right. & & 
 \nonumber \\
\left.
 {} +\quad \sum_{n=0}^{\infty} \epsilon^{2n}(n+1)
e^{-nv/2} V_{n}(\Theta)\right), & &
\end{eqnarray}

\begin{eqnarray*}
-\Theta'(1+\epsilon^{2} e^{-v/2} \sin^{2}\Theta) \quad = \quad  S_{\Theta} \quad 
= {} & &{} 
	\nonumber \\
{}  = \quad
e^{(u-v)/2} \sum_{n=0}^{\infty}\epsilon^{2n} e^{-nv/2}
V_{n}'(\Theta). & & {}
\end{eqnarray*}

The system (31) should be complemented by the smooth matching condition
at the point $x=x_{0},$  where  $\Theta(x_{0})= \pi/2 $ :

\begin{equation}
\sum_{n=1}^{\infty} \epsilon^{2n} e^{-nv(x_{0})/2} \left(V_{n}^{+'}(\pi /2) -
V_{n}^{-'}(\pi/2) \right) = 0.
\end{equation}
In particular, within the confines of $n=2$  approximation the Eq. (32)
allows us to find, through the use of (28) and (30), the approximate value of
the effective expansion parameter
\begin{equation} \zeta = \frac{1}{\displaystyle 3}
\epsilon^2e^{-v(x_0)/2}  = \frac{2}{\displaystyle 16\ln2-37/15} \approx 0.232.
\end{equation}

Now we can use (33) to estimate the mass length density  of the vortex
in search defined as
\begin{equation}
M=-\frac{2\pi}{\lambda^2} \Delta S,
\end{equation}
where $\Delta S$  denotes the total variation of $ S(x) $  in the domains $D^{\pm}$  :
\begin{eqnarray}
\arraycolsep = 1pt
\Delta S  =  S^{+}(\infty)  - S^{+}(x_{0}) +S^{-}(x_{0}) -
S^{-}(-\infty) = {} & &{}
   \nonumber \\
 {}= - 2 -\frac{1}{\nu}
e^{\frac{(n-v)}{2}}\sum_{n=1}^{\infty}\epsilon^{2n}
e^{-\frac{nv}{2}} (S_{n}^{+}-S_{n}^{-})|_{x=x_{0}} & &{}
\end{eqnarray}

Using (31) one gets in the first approximation in $\nu$ that
\begin{equation}
u=v=4x +\mathcal O(\nu).
\end{equation}
Inserting (36) into (35) and taking into account of (28) and (30), one
finds up to $\zeta^{2}$  terms that
\begin{equation}
\Delta S \approx - 2 - 4\zeta - 2\zeta^{2} \left(\frac{41}{15} - 8 \ln 2\right).
\end{equation}
Substituting (37) into (34) one gets, in view of (33), the approximate value
of the vortex mass
\begin{equation}
M \approx \frac{4\pi}{\lambda^{2}}1.31.
\end{equation}

As for the structure of the function $\Theta(x)$  defining the radial distribution
of matter inside the vortex,as well as those of the functions $u(x)$
and $v(x)$ describing the
gravitational field, they can be derived from the system (31) through the use
of the standard iteration technique.

\section{Discussion}

In conclusion it is worthwhile to underline that we considered here the case
of the neutral vortex not endowed with the topological charge density.
 Therefore after closing the vortex piece we come to the configuration with
 zero baryon number $Q$, this fact ensuing from the formula for the winding
 number
 \begin{equation}
Q =-\frac {1}{24\pi^2}\,\varepsilon^{ijk}\,\int  d^{3}x\,{\mathrm
 Sp}(l_i\,l_j\,l_k).
 \end{equation}
However, this shortcoming can be cured by generalizing the substitution (3)
with the aim to generate the topological charge density in the vortex. To
this end one can try the following ansatz:  \begin{equation}
U=\cos\psi\,\exp\,(i\sigma_{3}\chi) +
i\sigma_{1}\,\sin\psi\,\exp\,(i\sigma_{3}\delta),
\end{equation}
where $\sigma_{1},\,\,\sigma_{3}$  are the Pauli matrices and
$\psi,\, \chi,\,\delta$  stand for
the chiral angles represented in the cylindrical coordinates
$\rho,\,\,\varphi,\,\,z$ as follows:
\begin{equation}
\psi=\psi(\rho),\,\;\chi=-kz\eta(a-\rho),\,\;\delta=m\varphi,
 \end{equation}
with $ k = const,\,\, m\in{\mathcal Z}.$ The radius $a$ , entering
the argument of the Heaviside step function $\eta$, is chosen to satisfy the
piece-wise smooth matching condition $\cos\psi(a) =0$ . In the simplest case
we can impose the following boundary conditions:
\begin{equation}
\psi(0)=\pi,\, \;\psi(a)=\frac{\pi}{2},\,\;\psi(\infty)=0,
\end{equation}
and also the closure one:
\begin{equation}
k\,l=2 \pi n, \, \; n\in{\mathcal Z}, \, \; z\in[-\frac{l}{2},\frac{l}{2}],
\end{equation}
where $l$ stands for the length of the vortex piece.  Inserting (40)
into (39) yields, in virtue of (41), (42) and (43), the nontrivial baryon
number
\begin{equation}
Q = \frac{1}{4\pi^{2}}\int d^{3}x \sin 2\psi\left([\nabla\psi
\nabla\chi]\nabla\delta\right) = mn.
\end{equation} Thus,
the suggested substitution meets all necessary requirements.  The analysis of
the Skyrme - Einstein vortex configurations endowed with the topological
charge density will be exhibited in the subsequent paper.

 \end{document}